\documentclass[aps,pra,twocolumn,longbibliography,footinbib]{revtex4-1}
\usepackage{graphicx}
\usepackage{amsmath}
\usepackage{amssymb}

\newcommand {\scat} {\mathrm{sc}}
\newcommand  {\Ein}   {\mathrm{Ein}}

\begin{document}

\title{Optical-depth scaling of light scattering from a dense and cold atomic $^{87}$Rb gas}

\author{K.J. Kemp, S.J. Roof, and M.D. Havey}
\affiliation{Old Dominion University, Department of Physics, Norfolk, Virginia 23529}

\author{I.M. Sokolov${}^{1,2}$ and D.V. Kupriyanov${}^{1}$}
\affiliation{$^{1}$Department of Theoretical Physics, State Polytechnic
University, 195251, St.-Petersburg, Russia \\
$^{2}$Institute for Analytical Instrumentation, Russian Academy of Sciences, 198103, St.-Petersburg, Russia}

\author{W. Guerin}
\affiliation{Universit\'{e} C\^{o}te d'Azur, CNRS, Institut de Physique de Nice, France}

\date{\today}

\begin{abstract}
We report investigation of near-resonance light scattering from a cold and dense atomic gas of $^{87}$Rb atoms. Measurements are made for probe frequencies tuned near the $F=2\to F'=3$ nearly closed hyperfine transition, with particular attention paid to the dependence of  the scattered light intensity on detuning from resonance, the number of atoms in the sample, and atomic sample size. We find that, over a wide range of experimental variables, the optical depth of the atomic sample serves as an effective single scaling parameter which describes well all the experimental data.\\
\end{abstract}

%\pacs{42.50.Ct,42.25.Fx,42.50.Nn,32.70.Jz,32.80.Qk} % pacs are no longer in use by PRA

\maketitle

\section{Introduction}

Study of light interacting with cold and ultracold atomic gases is an active area of experimental and theoretical research\,\cite{Guerin:2017a,Kupriyanov:2017}. The subject appears to be deceptively simple, corresponding in many cases to a single weak probe beam scattering from a small cloud of cold atoms.  However, under most realistic situations, the atoms in such a sample interact not only with the incident radiation field, but also with the light scattered by all the other atoms in the sample.  The ensembles can then be viewed as many-body physical systems, and can display emergent complexity.  The optical response due to interaction with the near-resonance radiation reveals a collective optical response that differs significantly from that of a dilute and optically thin atomic ensemble.

As an illustration, Dicke's seminal paper~\cite{Dicke:1954} stimulated many theoretical and experimental studies on super- and subradiance.  These sudies focussed mostly on how a collection of excited atoms decay~\cite{Feld:1980,Gross:1982,Pavolini:1985}. More recently, the concept of superradiance has been extended to the single-photon regime~\cite{Scully:2006,Scully:2009}, and it has been shown that the problem is equivalent to that of classical dipoles driven by a weak field (linear-optics regime)~\cite{Svidzinsky:2010}. This triggered several experiments on the temporal decay dynamics of light scattered by cold atoms interacting with a weak probe beam, which allowed the observation of subradiance~\cite{Guerin:2016a} and superradiance~\cite{Araujo:2016,Roof:2016} in this regime. Note that nontrivial decay dynamics can also be due to diffuse scattering~\cite{Fioretti:1998,Labeyrie:2003,Weiss:2018}. Coherent transients in forward scattering give also rise to collective effects, such as a fast superflash~\cite{Kwong:2015}.

Steady-state experiments also reveal interesting collective effects, such as lensing~\cite{Roof:2015}, light diffusion~\cite{Labeyrie:2004,Saint-Jalm:2018}, changes in the radiation pressure force~\cite{Bienaime:2010,Bender:2010,Chabe:2014}, etc. Because of potentially important consequences for clock technology~\cite{Chang:2004}, the question of collective shifts of the resonance line, in particular, raised a lot of discussions~\cite{Friedberg:1973,Scully:2009b,Scully:2010,Manassah:2012,Javanainen:2014,Javanainen:2016} and experiments~\cite{Rohlsberger:2010,Keaveney:2012,Okaba:2014,Meir:2014,Bromley:2016,Jennewein:2016,Roof:2016,Corman:2017,Jennewein:2018}.
Even without any shift, changes in the line shape, collective broadening and saturation of the amount of scattered light have been observed in several experiments with different parameters and geometries, and interpreted somewhat differently~\cite{Labeyrie:2004,Balik:2013, Balik:2014,Pellegrino:2014,Jennewein:2016,Corman:2017,Jennewein:2018}.

In this context, we have previously reported measurements of light scattering from a high density and cold thermal gas of $^{87}$Rb\,\cite{Balik:2013, Balik:2014}.  These experiments, performed on the $F=2\to F'=3$ and $F=1\to F'=0$ nearly closed hyperfine transitions, focused on the time-dependent spectral development of light scattered from the atomic sample.  In those experiments it was observed that the measured intensity of the scattered light decreased with decreasing sample size containing a fixed number of atoms.  This observation suggested that collective effects may be important as the size of the atomic sample is changed.  In the present study, we have experimentally explored this question and report extensive measurements of the intensity of scattered light as a function of accessible parameters, including detuning from resonance excitation, atomic sample size, and the number of atoms contained in the sample at fixed sample size.  These measurements reveal evidence of the emergence of collective light scattering as a function of the experimental variables.   The measurements are found to be in good agreement with microscopic fully quantum calculations of the light scattering processes.  We find also that over a wide range of optical depths the experimental data is well described by a random walk simulation of light transport in the atomic medium; in this model the optical depth serves as an effective single scaling parameter.

In the following sections we first describe the measurement scheme and the experimental arrangement.  This is followed by presentation of the experimental results and comparison with quantum microscopic calculations. We follow this by a description of our random walk simulations, and comparison of the simulations with the peak optical depth dependence of the experimental data. Summary of the quantum microscopic calculation approach and other supporting information is deferred to the appendices.

\section{Experimental arrangement}

The basic experimental scheme has been described in detail elsewhere \cite{Olave:2015}; here we provide only an outline of details necessary to understand the experimental approach and results.  In the basic approach, we follow a multistep process to produce cold atom samples confined by a far off resonance trap (FORT). Initially, $^{87}$Rb  atoms are loaded into a 3-dimensional magneto-optical trap (MOT), with a density distribution that can be well described by a Gaussian distribution. The MOT is characterized using methods similar to those in \cite{Olave:2015}. The physical size and temperature of the MOT are found by directly measuring the radius of a fluorescence image projected onto a CCD camera (pixel resolution of 24 $\mu$m x 24 $\mu$m). The number of atoms trapped in the MOT is measured through traditional absorption imaging. The number is independently measured by using an optical pumping approach, as described in \cite{Chen:2001b}. We find that typically we have about 450 million atoms contained in the MOT.

A small fraction of the MOT atoms is loaded into the FORT. The trap consists of a single laser beam ($\lambda = 1064$ nm) focused to a transverse radius of about 20 $\mu$m. The intensity gradient of the focused light, along with being far detuned from resonance, creates a potential well in the ground state in which the atoms can be trapped. During the loading process, the MOT trapping laser is detuned $\sim10\gamma$ below resonance and the repumping laser is attenuated by $\sim99\%$. This reduces the radiation pressure and creates a compressed MOT, which has a better spatial overlap with the FORT laser beam. Atoms excited with the MOT trapping laser tuned near the $F=2 \to F^\prime=2$ transition undergo  inelastic Raman transitions, resulting in loading into the lower $F=1$ ground level. After a loading time of 70 ms, the trapping and repumping lasers are fully extinguished, along with the external magnetic field. Starting with an initial load of $1.3(2)\times 10^6$ atoms, the FORT laser is kept on for a minimum of 200 ms, until the sample is thermalized with $7.8(1)\times 10^5$ atoms at a temperature of about 100 $\mu$K. The FORT density distribution $\rho$ is well described by a Gaussian distribution as $\rho=\rho_0$exp${(-\frac{r^2}{2r_0^2}-\frac{y^2}{2y_0^2})}$ with a radial size $r_0$, longitudinal radius $y_0$, and peak density $\rho_0$.

\begin{figure}
\includegraphics[width=\columnwidth, keepaspectratio]{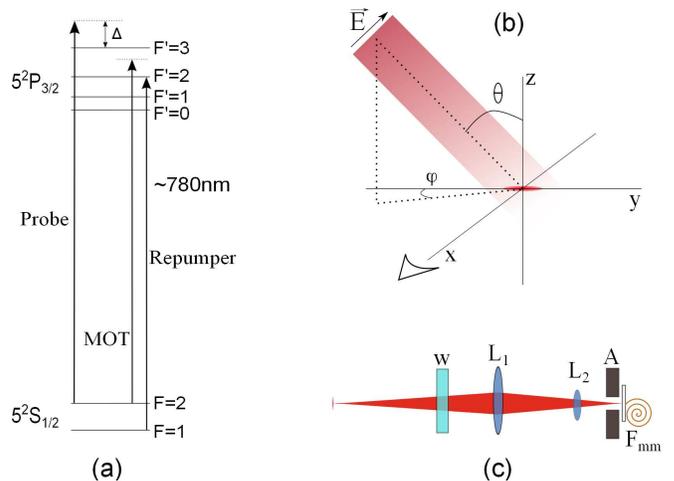}
\caption{The basic experimental scheme. (a) Relevant $^{87}$Rb energy levels.  (b) Geometry of probe optical excitation and fluorescence collection.  (c) Fluorescence detection arm.  Light is collected in the far field through a window (w) and focused into a 600 $\mu$m diameter multimode optical fiber $(F_{mm})$ with a pair of lenses $(L_1$ and $L_2)$ as shown.  }
\label{figure1}
\end{figure}

Once the atomic sample is thermalized, the FORT trapping laser is turned off. Initially the atoms are repumped into the $F=2$ ground state to prepare for probing on the $F=2 \to F^\prime=3$ transition. After an optical pumping phase of about 8 $\mu$s, nearly all of the atoms are transferred to the $F=2$ level. After another 2 $\mu$s, a near-resonance low intensity $(0.1 \ I_{sat})$ probe laser is flashed for 1 $\mu$s. The probe is offset from resonance by a detuning $\Delta$ = $f - f_0$, where $f_0$ is the bare atomic resonance frequency.  As shown in Fig \ref{figure1}, the probe beam is spatially much larger than the atomic sample, with a $e^{-2}$ radius of 4.5 mm, and is incident upon the sample at an oblique angle. The sample is allowed to continue to expand and is probed again 40 $\mu$s after the initial flash. This process continues for a total of 10 probe pulses up to a total expansion time of 370 $\mu$s. The sample expands from an initial volume with radii $r_0 = 3.0$ $\mu$m and $y_0 = 259$ $\mu$m to final radii of $r_0 = 33.4$ $\mu$m and $y_0 = 261$ $\mu$m. The fluorescence from the sample is collected during all 10 pulses and focused into a multimode fiber connected to an infrared sensitive photomultiplier tube (PMT). The output of the PMT is directed without amplification to a multichannel scaler having 40 ns time resolution.  For the results presented in this paper, this time signal was integrated over the duration of each individual pulse to show the total amount of fluorescence for each sample size, all while maintaining the same number of atoms.

\begin{figure}
\includegraphics[width=\columnwidth, keepaspectratio]{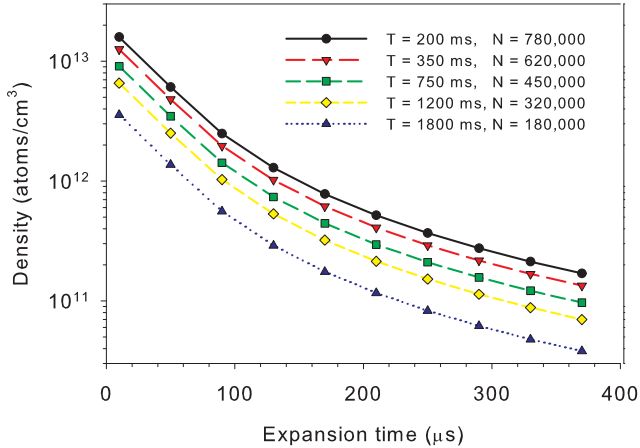}
\caption{Reduction of the number of atoms within the FORT over time due to ballistic expansion, thermalization and background gas collisions. After various hold times T there are N atoms in the trap (see legend).  After the FORT trapping laser is extinguished, the sample expands, reducing the peak atomic density as shown. }
\label{figure2}
\end{figure}

In order to sample a broader range of atomic sizes and densities, the number of atoms can also be changed. The peak density of the sample depends on the holding time of the FORT; background gas collisions decrease the number of atoms within the sample. At the longest hold time used for these measurements (2.5 s), the number of atoms is reduced to $1.8(7)\times 10^5$. In Fig \ref{figure2}, the peak density for each sample holding time as a function of expansion time is shown. Finally, we also studied the dependence of the scattered light intensity on probe detuning at the highest possible density for our thermalized sample. Using an acousto-optical modulator (AOM) in a double-pass setup, the frequency of the probe laser was tuned over a range of nearly 60 MHz while maintaining a constant probe optical power.

\section{Results and discussion}

In this section we present our experimental results and make side by side comparison of the measurements and fully quantum calculations of the measured quantities.  The details of the calculational techniques are described in detail in several earlier papers \cite{Sokolov:2009, Sokolov:2011,Kupriyanov:2017} on the general subject of light scattering in a cold and dense gas.  Our approach is also sketched in Appendix A of this paper.  Note that the theoretical results are scaled \cite{Sokolov:2013} to account for the fact that the measurements and theoretical comparisons are made at very different numbers of atoms.  These results and comparisons are followed by two subsections in which the data is globally analyzed and discussed in terms of attenuation of the propagating light beam and a random walk for the diffusing light.

\subsection{Experimental results and comparison with theory}

We first point out that, in all cases, fluorescence measurements are made after the atoms in the FORT have essentially thermalized and the FORT has been turned off, so that the atoms are mainly in free space. There are two primary overlapping experimental protocols.  In one, once the FORT has been extinguished, the expanding atomic sample is exposed to a series of ten 1 $\mu s$ probe pulses temporally spaced to map out a factor of several hundred in peak atomic density.  As the probe spatial profile is much larger than the atomic sample, the number of atoms probed remains essentially constant.  In a second protocol, the atom sample is held in the trap for increasingly longer periods of time; background gas collisions reduce the number of atoms in the ensemble, while the sample size, as measured by the sample Gaussian radii, remains the same.   Then the FORT is extinguished and a sequence of probe pulses is used to probe the sample.   This dual approach allows mapping out of both the atomic sample size and atomic density dependence of the fluorescence signals.

\begin{figure}
{\includegraphics{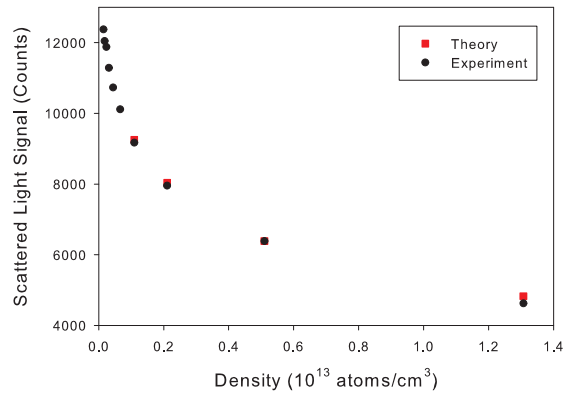}}
\caption{On resonance variation of the scattered light signals with peak atomic density.  Note the strong increase of the signal size with decreasing atomic density, for a fixed number of atoms in the sample.}
\label{figure3}
\end{figure}

As an initial result, we present in Fig. 3 the measured fluorescence signals from a 10 $\mu s$ probe pulse and their dependence on the peak atomic density.  We see in the figure that the signals increase with decreasing atomic density.  The origin of this somewhat counterintuitive effect arises from the fact that, for the highest densities, and consequently the greatest optical depth, the probe beam is attenuated during its traversal through the sample.  The scattering signals then should originate mainly from light scattered from the illuminated outer regions of the sample surface, and the relatively fewer atoms compared to the sample as a whole. We will study in more detail this ``shadow effect'' in the next subsection. As the density is decreased, on the other hand, the sample becomes more optically thin; the sample ultimately scatters light as a collection of individual atoms. Comparison of the experimental results with calculations shows very good agreement.  Note that the vertical (signal) scale is adjusted to match the experimental and theoretical responses.

\begin{figure}\scalebox{1.2}
{\includegraphics[width=\columnwidth, keepaspectratio]{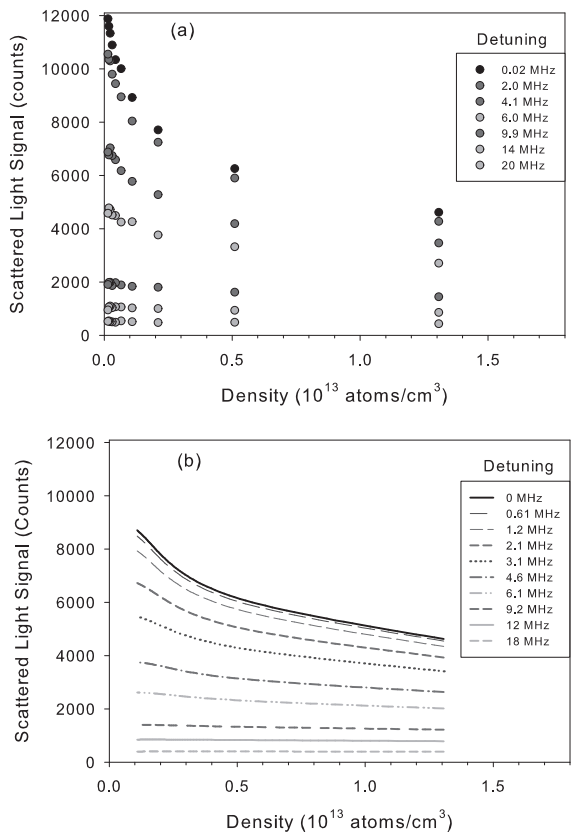}}
\caption{Detuning and density dependence of the measured scattered light intensity. (a) Experimental results for positive (blue) detunings. (b) Theoretical results. The vertical scale has been adjusted to match the experimental data.}
\label{figure4}
\end{figure}

We elaborated on this general effect by measuring the dependence of the scattering signals on atomic density and on detuning from atomic resonance.  The overall experimental results for all positive blue detunings and densities are shown in Fig. 4(a).  One striking feature of these results is that, for larger detunings, the sensitivity of the signals to decreases in the density is significantly reduced, and for the largest detunings from resonance, there is, within the experimental uncertainty, no variation of the signal intensity with peak atomic density.  This effect is due to the decreasing optical depth of the atomic sample with increasing detuning; for the smallest optical depth, all the atoms experience essentially the same probe intensity, and thus contribute to the scattering signals. The corresponding theoretical results are shown Fig. 4(b).  These results are in very good qualitative agreement with the experimental ones.  Red detuned measurements (not shown) are also in very good agreement with the simulations.  The data are also quite symmetric about zero detuning; this is seen in the  characteristic spectral response for two different densities, as shown in Fig. 5.  There the solid lines represent Lorentzian spectral profiles; this line shape is a very good empirical fit to the measured profile.

\begin{figure}
{\includegraphics{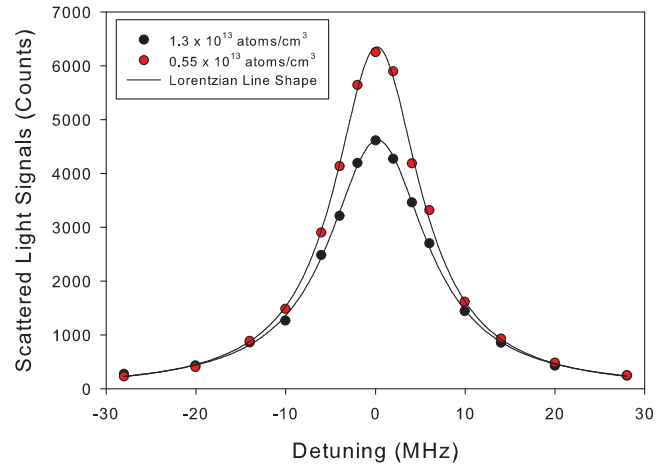}}
\caption{Representative line shapes for the dependence of the measured signals on detuning from atomic resonance.}
\label{figure5}
\end{figure}

Implicit in Figs. 3 and Fig. 4 is a dependence on the spectral width ($viz.$ Fig. 5) and the ensemble response to changes in atomic density.   This dependence is shown in Fig. 6, where we see a nonlinear increase of the spectral width with increasing density.  This dependence is qualitatively due to the fact that major contributions to the signal arise from atoms near the outer regions of the atomic sample, the deeper atoms contributing less due to the so-called shadow effect.  For a large optical depth and a uniform density, this implies a roughly $\sqrt{b}$ scaling of the width; here $b$ is the peak optical depth through the center of the sample.  Realistically, our samples are strongly inhomogeneous, and there are contributions to the signals from a range of atomic densities.  Such scaling should then be considered only as a qualitative feature of the measured spectral widths.

Finally, we have examined the dependence of the measured scattered light intensity with variations in the effective volume of the sample.  We use as a measure of the sample volume the product of the atom sample Gaussian radii, $\emph{viz.}$, $(2 \pi)^{3/2} y_o r_{o}^{2}$.  In these measurements, this product is held fixed as the number of atoms in the sample is varied. Results are shown in Fig. 7.  We see in Fig. 7 that, for each sample size, and within the experimental uncertainty, the signal increases monotonically with increasing number of atoms (or atomic density). However, the rate of increase is significantly different, depending on the sample size, and is strongly suppressed for the smallest sample sizes.

\begin{figure}
\scalebox{0.8}
{\includegraphics{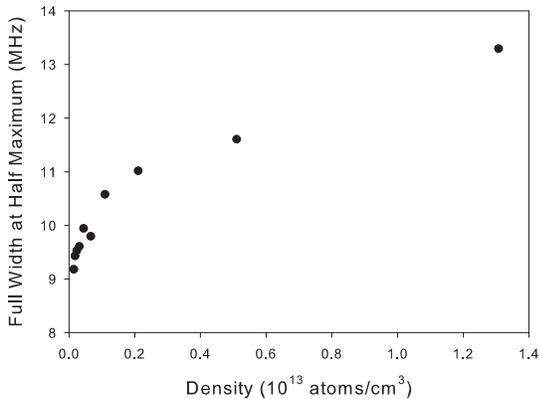}}
\caption{Dependence of the full width at half maximum of the atomic resonance response as a function of atomic density.  These measurements correspond to varying the density by changing the sample size while holding the number of atoms fixed.  }
\label{figure6}
\end{figure}

\begin{figure}
\scalebox{0.7}
{\includegraphics{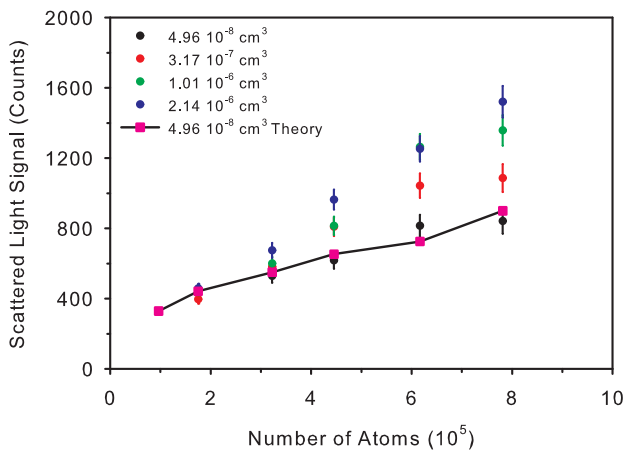}}
\caption{Representative atom number dependence of the scattering signals as a function of the cold atom sample size.  The data is labeled according to the volume of the sample, as described in the text.}
\label{figure7}
\end{figure}

\subsection{Rescaling according to the Beer-Lambert Law}

The good agreement between the data and the full microscopic theory is in itself satisfactory but it does not allow identifying the relevant physical ingredients at the origin of the specific behavior of the scattered light as a function of the different control parameters. This is because the microscopic theory includes many effects: attenuation of the probe light, diffraction and refraction, multiple scattering, super and subradiance, collective shifts, etc. It is thus useful to compare the data with a much simplified theory, including only some of these effects.

An effective approximation could be based on the ladder-type expansion of the light correlation function, which leads to a Bethe-Salpeter type equation.  This can be numerically solved via a sequence of iterative steps (multiple scattering events), see \cite{Kupriyanov:2017}.  Such an approach evidently ignores any cross interference in the process of multiple scattering, which seems a rather realistic assumption for a dilute and disordered atomic gas.  The applicability of the Bethe-Salpeter approach has been successfully demonstrated for the theory of random lasing, see \cite{Gerasimov:2014,Gerasimov:2015}

In this section we show that even in taking a simpler approximation by transforming the Bethe-Salpeter equation to the light transport equation, and
taking into account only the attenuation of the probe beam in the atomic sample, following the Beer-Lambert law, is enough to explain the data with a rather good agreement. This shows that the main physical ingredient of the experiment is the so-called ``shadow effect'': atoms at the back of the sample are less illuminated by the incident laser, which induces an effective reduction of the total scattering cross-section compared to a collection of independent atoms illuminated by the same laser intensity. As explained in detail in~\cite{Bachelard:2016}, this effect also explains previous observations of a collective reduction of the radiation pressure force~\cite{Bienaime:2010,Chabe:2014}. It could also explain the results of~\cite{Pellegrino:2014}, although the very small sample sizes and high densities used in that work might induce some other effects.

From the Beer-Lambert Law, one can easily show (see Appendix\,\ref{App.shadow} or ref.\,\cite{Chabe:2014}) that the total scattering cross-section of a Gaussian cloud (containing $N$ atoms and illuminated by a plane wave) is
\begin{equation}\label{eq.Sigma_RW}
\Sigma_\scat = N \sigma_\scat \times \frac{\Ein(b)}{b} \, ,
\end{equation}
where Ein is the integer function~\cite{Wolfram:Ein}
\begin{equation}
\begin{split}
\Ein(b) & = \int_0^b \frac{1-e^{-x}}{x} \, dx \\
& = b \left[ 1 + \sum_{n=1}^\infty \frac{(-b)^n}{(n+1)(n+1)!} \right] \,,
\end{split}
\end{equation}
and $\sigma_\scat$ is the single-atom scattering cross-section.
Here $b$ is the optical depth along the line of sight and the factor $\Ein(b)/b$ in Eq.\,\ref{eq.Sigma_RW} corresponds to the deviation from single-atom physics induced by the shadow effect.
In the limit of vanishing optical depth $b$, the value expected from single atom physics is recovered, $\Sigma_\scat=N\sigma_\scat$. For high optical depth, the cross-section increases only logarithmically, which appears as a collective saturation of the scattered light.

Let us now use this result to rescale the experimental data. The measured scattered light is proportional to $\Sigma_\scat$. For data taken with a fixed atom number and varying detuning (``protocol 1''), such as the data reported in Fig.\,\ref{figure4}(a) and Fig.\,\ref{figure5}, one should divide the signal by $\sigma_\scat \propto 1/(1+4\Delta^2/\Gamma^2)$ and compare the results to $\Ein(b)/b$. For data acquired at a fixed detuning and varying atom number (``protocol 2''), such as the data reported in Fig.\,\ref{figure7}, one should divide the signal by $N$ and also compare to $\Ein(b)/b$. In both cases one has to allow a global multiplicative factor to fit the data to the theoretical curve, since the signal is not calibrated in absolute value. The relevant optical depth $b$ is the one along the line of sight of the laser, given by
\begin{equation}
b = \frac{\sqrt{2\pi}\rho_0\sigma_\scat r_0}{\sqrt{ \cos^2\theta + \sin^2\theta \sin^2 \phi + \eta^2 \sin^2 \theta \cos^2\phi }} \, ,
\end{equation}
where $\eta = r_0/y_0$, $r_0$, $y_0$, $\rho_0$ vary during the expansion and the angles $\theta,\phi$ are given by the geometry of the experiment as shown in Fig.\,\ref{figure1} ($\theta=23^\circ$ and $\phi = 30^\circ$).

\begin{figure}
\includegraphics{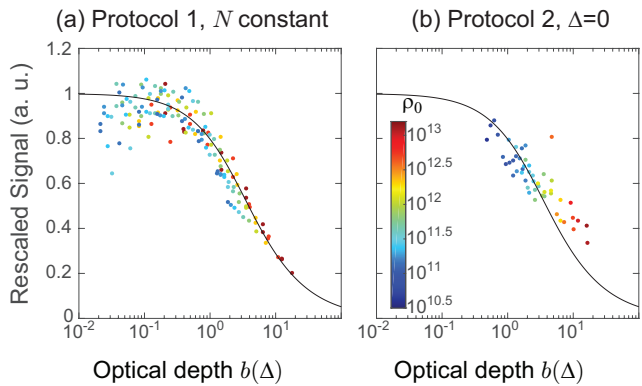}
\caption{Rescaled experimental data: the light scattering signal is plotted as a function of the detuning-depended optical depth $b(\Delta)$ and the color code indicates the peak density $\rho_0$ in cm$^{-3}$ (log scale). The solid line is the function $\Ein(b)/b$ which describes the shadow effect from the Beer-Lambert Law. The two panels correspond to the two different experimental protocols, the first one with a varying detuning and a constant atom number, the second one with the laser on resonance and a varying atom number. In both cases the sizes of the cloud also vary, and thus the volume and density. A global vertical scaling factor for each data set is the only free parameter.}
\label{figure8}
\end{figure}

We show the rescaled data in Fig.\,\ref{figure8}. The two panels correspond to the two different experimental protocols. The striking result is that, despite the different protocols and different orders of magnitudes (almost 3 orders of magnitude in density and in optical depth), all data points collapse quite close to the curve $\Ein(b)/b$ describing the shadow effect, demonstrating that it is indeed the main physical ingredient of the collective behavior of the scattered light intensity.

\subsection{Impact of multiple scattering}

The previous scaling based on the total scattering cross-section supposes that the light is emitted isotropically from the atomic sample. This is not the case when the optical depth is large, as already studied in~\cite{Labeyrie:2004}, although the anisotropy is much less pronounced when the cloud is illuminated by a wide beam (plane wave), as is the case here, compared to the case when a large cloud is illuminated by a narrow beam, as in~\cite{Labeyrie:2004}.

To describe this effect one needs to take into account multiple scattering of light inside the sample. This is naturally included in the microscopic model, but it is also possible to use stochastic simulations based on a random walk algorithm for light. In such a model,  cooperative and coherent effects such as super and subradiance, interference and diffraction are neglected, but one can well describe diffuse light scattering with the true parameters of the experiments (also including subtle effects like the frequency redistribution due to Doppler broadening, if needed, see e.g.~\cite{Labeyrie:2003,Eloy:2018,Weiss:2018}).

\begin{figure}
\includegraphics{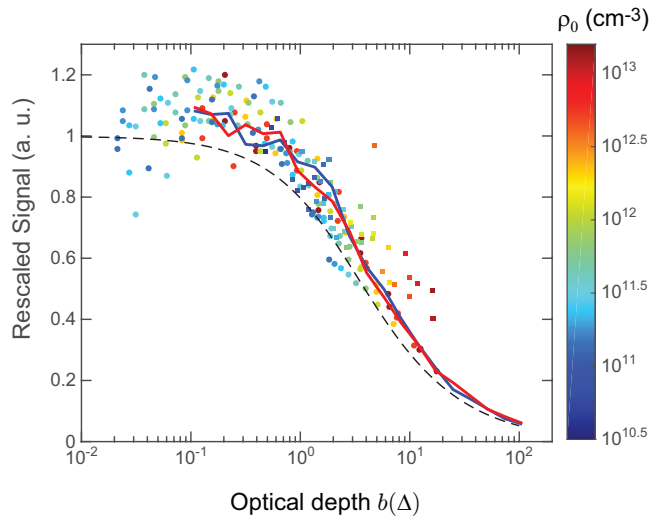}
\caption{Comparison between the experimental data and the random walk simulation. The data are rescaled like in Fig.\,\ref{figure8}, dots corresponds to the protocol 1 and squares to the protocol 2. The color code indicates the peak density $\rho_0$ in cm$^{-3}$ (log scale). The solid lines are the results of the random walk simulations for the two extreme aspect ratios of the cloud, $\eta = r_0/y_0 \simeq 0.13$ (blue) and $\eta \simeq 0.013$ (red). The dashed line is the function $\Ein(b)/b$ which describes the shadow effect from the Beer-Lambert Law. The global vertical scaling factor for each data set has been adapted to match the random walk results.}
\label{figure9}
\end{figure}

We have performed such random walk simulations for varying optical depths. The simulations include the actual geometry of the laser beam (size and direction) and of the detection (direction), the anisotropy of the scattering diagram for the first scattering event and the Gaussian density distribution of the cloud. We use the size $y_0$ of the cloud, which is almost constant for all data points, and the two extreme transverse sizes, corresponding to the shortest and longest time of flight.  We do not take into account the Doppler-induced frequency redistribution during multiple scattering as it should be a tiny effect with the moderate temperature and optical depths explored here. The results are shown in Fig.~\,\ref{figure9}.

The comparison between the random walk simulations and the simple Beer-Lambert prediction shows a small difference: the scattered light signal is always slightly larger in the random walk simulations. Several contributions explain this difference. First, the Gaussian beam profile has a stronger intensity at the center, where it interacts with the cloud, compared with a plane-wave illumination. Second, the small anisotropy of the scattering diagram of Rb (we suppose an equipopulated mixture of Zeeman states) slightly favors the direction of detection. And third, at large optical depths, multiple scattering takes place and light has a higher probability to escape along the backward and transverse directions, which also favors the detection direction compared to an isotropic emission. Finally, at the precision of the numerical simulation, we do not see any significant difference between the two extreme aspect ratios of the cloud, showing that this parameter does not affect the results.

In Fig.\,\ref{figure9} the vertical scaling factor of each data set has been chosen to match the simulation results. With this as the only free parameter the simulations and the experimental points are in very good agreement.

\section{Conclusions}

Using two different experimental protocols, we have made measurements of diffusive light scattering from a cold thermal gas of $^{87}$Rb.  Due to variations in the number of atoms in the sample, or the size of the sample at fixed number of atoms, the experiments extended over almost 3 orders of magnitude in density and in optical depth.  The measured diffusive light spectra were found to be in very good agreement with fully quantum based calculations.  A second and simpler analysis approach used stochastic simulations based on a random walk algorithm for the multiply scattered light. The simulations revealed that the optical depth of the atomic sample can serve as an effective single scaling parameter which describes very well all the experimental data.

\section*{Acknowledgments}

We appreciate financial support by the National Science Foundation (Grant No. NSF-PHY-1606743), the Russian Foundation for Basic Research  (Grant No. 18-02-00265), the Russian Scientific Foundation (Grant No. 18-72-10039), the charitable foundation BASIS (Grant No. 18-1-1-48-1), and the French Agence Nationale pour la Recherche (project LOVE, No. ANR-14-CE26-0032). The quantum microscopic calculations were financially supported by the Russian Science Foundation under Project 17-12-01085.

\appendix

\section{Quantum microscopic approach.}\label{App.theory}

In this section we provide a sketch of the quantum microscopic method we have
used as one of the possible approaches for the theoretical description of light scattering discussed in this paper. In this sketch, we follow the descriptions in earlier papers \cite{Sokolov:2011,Balik:2013,Kuraptsev:2017,Kupriyanov:2017}.

In addition, numerical simulations using the actual rather large number of atoms contained in the experimental samples are very difficult.  In order to compare the theoretical results with experimental ones, the calculated results are scaled to the proper number of atoms using an approach developed earlier by us \cite{Sokolov:2013}.

In our calculations of time-dependent fluorescence we  solve the nonstationary Schr\"odinger equation for the wave function $\psi$ of the joint system consisting of all atoms and a weak electromagnetic field. A vacuum reservoir is also included in our considerations.

We consider a disordered atomic cloud of $N$ identical motionless two-level atoms with a ground state $J_g = 1$ separated by the frequency $\omega_a$ from the excited state $J_e = 1$. The decay constant of this state is $\gamma$. Atoms are assumed to be located at random positions $\mathbf{r}_i,\,  (i = 1, . . . ,N)$. Possible atomic displacement caused by residual atomic motion is taken into account by averaging of calculated quantities over this random spatial distribution of the atoms.

We search for the wave function $\psi$ as an expansion in a set of eigenfunctions of the Hamiltonian $H_0$ of the noninteracting atoms and field. Taking into account that the exciting radiation is weak, we account only for states with no more than one photon in the field. For the Fourier component of the amplitude of these states we have a infinite set of equations because of the infinite number of states with one photon. We exclude amplitudes  of these  states  and obtain a finite closed system of equations for a one fold excited atomic system. The solution of this system gives us the opportunity to find all the other amplitudes and consequently the approximate wave function of the considered joint physical system.

Knowledge of the wave function allows us to describe the properties of the atomic ensemble as well as the properties of the secondary radiation. Particularly, the intensity $I_\alpha (\mathbf{\Omega},t )$ of the polarization component $\alpha$ of the light scattered in a unit solid angle around the direction given by the radius-vector $\mathbf{r}$ ($\mathbf{\Omega}={\theta,\varphi}$) can be determined as follows (for more details see \cite{Sokolov:2011})
\begin{equation} I_\alpha (\mathbf{\Omega},t )=\frac{c}{4\pi}  \left\langle \psi
\right\vert E^{(-)}_\alpha(\mathbf{r}) E^{(+)}_\alpha(\mathbf{r}) \left\vert \psi \right\rangle r^2. \label{A1}
\end{equation}
Here $E^{(\pm)}_\alpha(\mathbf{r})$ are the positive and negative frequency parts of the electric field operator.

As was shown in \cite{Sokolov:2011} the Fourier transform of the matrix element (\ref{A1}) is
\begin{eqnarray}
\left\langle \psi \right\vert E^{(-)}_\alpha(\mathbf{r}) E^{(+)}_\alpha(\mathbf{r}) \left\vert \psi \right\rangle = \left\vert\int\limits_{-\infty }^{\infty
}\dfrac{\hbar\exp(-i\omega t)d\omega }{2\pi } \right. \notag  \\
\left.  \sum\limits_{e,e^{\prime }}\widetilde{\Sigma }_{\alpha e}(\omega )R_{ee^{\prime }}(\omega )\Lambda_{e^{\prime }}(\omega ) \right\vert ^{2}.  \label{A2}
\end{eqnarray}
Here the vector $ \Lambda_{e}(\omega )$ describes atomic excitation by external radiation
\begin{equation} \Lambda_{e}(\omega)=-\frac{\mathbf{d}_{e;g}\mathbf{E}(\omega)}{\hbar }=
-\frac{\mathbf{u}\mathbf{d}_{e;g}}{\hbar }E_0(\omega)\exp(i\mathbf{kr_e}). \label{A3}
\end{equation}
In this equation $\mathbf{d}_{e;g}$ is the dipole matrix element for the transition from the ground $g$ to the excited $e$ state of the atom, $E_{0}(\omega)$ is a Fourier amplitude of the external radiation, which we assume to be a plane wave with  wave vector  $\mathbf{k}$ and unit polarization vector $\mathbf{u}$; $\mathbf{r}_{e}$ is the radius-vector of the atom excited in the state $e$.

The matrix $R_{ee^{\prime }}(\omega )$ is the resolvent of the considered system projected on the one-fold atomic excited states
\begin{eqnarray}
R_{ee^{\prime }}(\omega )=\left[ (\omega -\omega _{e})\delta _{ee^{\prime }}-\Sigma _{ee^{\prime }}(\omega)\right] ^{-1}.  \label{A4}
\end{eqnarray}
We determine it numerically on the basis of the known expression for the matrix $\Sigma _{ee^{\prime }}(\omega)$ found in \cite{Sokolov:2011}. Matrix elements $\Sigma _{ee^{\prime }}(\omega )$ for $e\neq e^\prime$ describe excitation exchange between different atoms
\begin{eqnarray}
\label{A5} &&\Sigma _{ee^{\prime }}(\omega )=\sum\limits_{\mu ,\nu} \frac{\mathbf{d}_{e_{a};g_{a}}^{\mu }\mathbf{d}_{g_{b};e_{b}}^{\nu }}{\hbar r^{3}}\times \\&& \left[ \delta
_{\mu \nu }\left( 1-i\frac{\omega _{a}r}{c}-\left( \frac{\omega _{a}r}{c}\right) ^{2}\right) \exp \left( i\frac{\omega _{a}r}{c}\right) \right. - \notag  \\&& \left.
-\dfrac{\mathbf{r}_{\mu }\mathbf{r}_{\nu }}{r^{2}}\left( 3-3i\frac{\omega _{a}r}{c}-\left( \frac{\omega _{a}r}{c}\right) ^{2}\right) \exp \left( i\frac{\omega _{a}r}{c}\right)
\right]. \notag
\end{eqnarray}
This expression is written within the framework of the pole approximation ($\Sigma _{ee^{\prime }}(\omega )=\Sigma _{ee^{\prime }}(\omega_a )$, see \cite{Sokolov:2011}). Besides that, we assume that in states $\psi _{e^{\prime }}$  and $\psi _e$ atoms $b$  and $a$ are excited correspondingly. In (\ref{A5}) $\mathbf{r}_\mu$ is projections of the vector $\mathbf{r}=\mathbf{r}_{a}-\mathbf{r}_{b}$ on the axes of the chosen reference frame and  $r=|\mathbf{r}|$ is the spacing between atoms  $a$ and $b$.

If $e$ and $e^\prime$ correspond to excited states of one atom then $\Sigma _{ee^{\prime }}(\omega )$ differs from zero only for $e=e^\prime$ (i.e. for $m=m^\prime$, where $m$ is magnetic quantum number of the atomic excited state). In this case
\begin{equation} \Sigma _{ee}(\omega )=-i\gamma/2.  \label{A6}
\end{equation}

The matrix $\widetilde{\Sigma }_{\alpha e}(\omega )$  in (\ref{A2}) describes light propagation from an atom  $e$ to the photodetector. In the rotating wave approximation it is (see \cite{Sokolov:2011})
\begin{eqnarray}
\label{A7} \widetilde{\Sigma }_{\alpha e}(\omega)=-\frac{\mathbf{u'}_{\alpha }^{\ast}\mathbf{d}_{g;e}}{\hbar r } \left( \frac{\omega }{c}\right) ^{2}\exp \left( i\frac{\omega
\left\vert\mathbf{r-r}_{e}\right\vert }{c}\right)  \\ \notag \approx
 -\frac{\mathbf{u'}_{\alpha }^{\ast}\mathbf{d}_{g;e}}{\hbar r } \left( \frac{\omega
}{c}\right) ^{2}\exp \left( i\frac{\omega r }{c}-i\frac{\mathbf{k'r}_e}{c}\right) .
\end{eqnarray}
Here $\mathbf{u'}_{\alpha }^{\ast}$ is a unit polarization vector of the scattered wave and $\mathbf{k'}$ is its wave vector.

Substituting  (\ref{A3}) and  (\ref{A7}) into  (\ref{A2}), after some simplifications we have
\begin{eqnarray}
\label{A8}
&&I_\alpha(\mathbf{\Omega},t )=\frac{c}{4\pi\hbar^2} \left\vert\int\limits_{-\infty }^{\infty }E_0(\omega)k^{2}\dfrac{\exp(-i\omega t)d\omega }{2\pi } \right. \\
&& \sum\limits_{e,e^{\prime }}\left.\left(\mathbf{u}^{\prime \ast }\mathbf{d}_{g;e}\right) R_{ee^{\prime }}(\omega ) \left(\mathbf{ud}_{e^{\prime };g}\right) \exp \left(
i(\mathbf{kr}_{e^{\prime }}-\mathbf{k}^{\prime}\mathbf{r}_{e})\right)  \right\vert ^{2}. \notag
\end{eqnarray}
The total intensity $I(\mathbf{\Omega},t) $ can be obtained as a sum of (\ref{A8}) over two orthogonal polarizations $\alpha$.

In the present work we calculate the integrals (\ref{A8}) by means of the residue theory. For this purpose we find the
poles of the matrix of the resolvent $R_{ee^{\prime }}$. These poles are determined in turn by eigenstates of the matrix $\Sigma _{ee^{\prime }}$. Decomposing the vector $ \Lambda_{e}(\omega )$ (\ref{A3}) over the eigenvector of the matrix $\Sigma _{ee^{\prime }}$ we present the integral (\ref{A8}) as the sum of separate poles contributions. The  decay constant and energy of each pole are determined by the imaginary and real part of eigenvalues of $\Sigma _{ee^{\prime }}$.

The described consequent quantum approach allowed us to consider atomic clouds with several thousands of atoms. As mentioned earlier, the samples explored experimentally contain several orders of magnitude more atoms. To have the possibility to compare theoretical results with performed experiments we use approximate scaling laws obtained previously \cite{Sokolov:2013}. Those scaling laws were obtained by exploring numerically the dependence of the differential scattering cross sections on sample size and give us an opportunity to predict the scattered light intensity for the experimentally studied atomic ensembles on the basis of results obtained for macroscopic  clouds.

Another problem in comparison of the theoretical and experimental results connects with the  difference in level structure of atoms $^{87}$Rb and that considered in theory, a $J=0 \rightarrow J=1$ transition.   We take into account this difference by proper choice of the atomic density. The key parameter which determine the nature of collective effects is mean free path of the photons. By this reason in calculations we choose the density
in such a way that photons would have the same mean free path as in the $^{87}$Rb samples. Estimating the resonant cross section of the light from a single atom with $J=0\rightarrow J=1$ transition as $3\lambda^2/2\pi$ we obtain that $n_0\simeq 5\cdot\,10^{13}\,cm^{-3}$ in experiment corresponds to $n_0\simeq 0.05\,k_0^{-3}$ in theory. Here $k_0$ is the wave number of the scattered light.

\section{Proof of Eq.\,(\ref{eq.Sigma_RW}) for the shadow effect}\label{App.shadow}

For simplicity, let us take an isotropic Gaussian cloud with density distribution $\rho = \rho_0 e^{-r^2/(2R^2)}$ and consider a plane wave (intensity $I_0$) propagating along $z$. The transmitted intensity has a transverse distribution
\begin{equation}
\begin{split}
I_T(\mathbf{r}_\perp) & = I_0 \exp\left(-\rho_0 \sigma_\scat \int e^{-r^2/(2R^2)} dz\right) \\
      & = I_0 \exp\left(-b e^{-r_\perp^2/(2R^2)}\right) \, ,
\end{split}
\end{equation}
with $b = \sqrt{2\pi}\rho_0\sigma_\scat R$ and $\mathbf{r}_\perp=(x,y)$.

Moreover, what is scattered is what is not transmitted, so we have
\begin{equation}
\Sigma_\scat =  \frac{P_\scat}{I_0} = \int \left[ 1- \exp\left(-b e^{-r_\perp^2/(2R^2)} \right) \right] d^2\mathbf{r}_\perp\,.
\end{equation}
Using $d^2\mathbf{r}_\perp = 2\pi r_\perp dr_\perp$ and the change of variable $u=be^{-r_\perp^2/(2R^2)}$ one obtains
\begin{equation} \label{eq.Sigma_RW_1}
\Sigma_\scat = 2\pi R^2 \int_0^{b} \frac{1-e^{-u}}{u} du
 = 2\pi R^2  \Ein(b) .
\end{equation}

For single atom physics, the total cross section would be $N\sigma_\scat$. Using $b = \sigma_\scat / (2\pi) \times N/R^2$, it is thus physically meaningful to write
\begin{equation} \label{eq.Sigma_RW_2}
\Sigma_\scat = N \sigma_\scat \times \frac{\Ein(b)}{b} \; ,
\end{equation}
which is Eq.~(\ref{eq.Sigma_RW}).

%\bibliography{AllMyBiblio}  % default style is apsrev4-1.bst, use option longbibliography (at the beginning) to have the titles
%

\end{document}